\documentclass[aps,prl,amssymb,amsmath,superscriptaddress,twocolumn]{revtex4-1}

\usepackage{color}
\usepackage{dsfont}
\usepackage{graphicx}
\usepackage{amsthm}
\usepackage{amsfonts}
\usepackage{braket}

\newcommand{\di}{i}

\begin{document}
\title{Algebraic maximum violation of the Leggett-Garg inequality in a two-level system \\ under 
$\mathcal{P}\mathcal{T}$ symmetric dynamics}
\author{H S Karthik}
\email{hskarthik@imsc.res.in}	
\affiliation{Optics and Quantum Information Group, The Institute of Mathematical Sciences, HBNI, C. I. T. Campus, Taramani, Chennai 600113, India}
\author{Akshata Shenoy Hejamadi}
\email{akshataphy@gmail.com} 
\affiliation{Optics and Quantum Information Group, The Institute of Mathematical Sciences, HBNI, C. I. T. Campus, Taramani, Chennai 600113, India}
\author{A R Usha Devi}
\email{arutth@rediffmail.com}
\affiliation{Department of Physics, Bangalore University, 
	Bangalore-560 056, India}
\affiliation{Inspire Institute Inc., Alexandria, Virginia, 22303, USA}
\date{\today}

\begin{abstract}
We investigate the three-term Leggett-Garg inequality (LGI) for a two-level quantum system undergoing \textit{parity-time} ($\mathcal{P}\mathcal{T}$) symmetric dynamics governed by a non-Hermitian Hamiltonian, when a sequence of dichotomic projective measurements are carried out at different time intervals. In contrast to the case of coherent unitary dynamics,  violation of LGI is shown to increase beyond the \textsl{temporal} Tsirelson bound approaching the algebraic maximum in the limit of the spontaneous $\mathcal{P}\mathcal{T}$ symmetry breaking. 
\end{abstract}

\maketitle

\textit{Introduction.}--
Quantum and classical descriptions of correlations between observables of spatially separated systems differ drastically. Bell-CHSH inequalities \cite{B64,CHS+69} place bounds on spatial correlations  based  on the classical framework of \textit{local realism}. However, quantum theory predicts the  presence of  \textit{non-local} correlations, which cry out for explanations~\cite{Bell1}. Experimental violation of the Bell-CHSH inequalities by quantum \textit{entangled} states project how quantum theory differs from our \textit{intuitive} classical description of correlations in spatially separated systems \cite{BCP+14}. Moreover, it has been realized that the Tsirelson bound~\cite{Tsirelson80}, the maximum possible violation of Bell-CHSH inequality in quantum theory is less than the algebraic maximum value. Violation of the Bell-CHSH inequality beyond the Tsirelson bound, points towards the existence of post quantum theories~\cite{PR92,PR94}, which attract  attention from the perspective of quantum information theory.

On the other hand, Leggett and Garg probed the contrast between the conflicting worldviews underlying the quantum and classical theories of nature in terms of \textit{temporal correlations} in a single system under the tenet of \textit{macrorealism}~\cite{LG85}. Macrorealism is based on the twin assumptions: (i) Macrorealism per se: Physical properties of a macroscopic system exists independent of the act of observation. (ii) Non-invasive measurability: Measurements performed do not influence the subsequent system evolution. Leggett-Garg inequality~(LGI) places macrorealistic  bounds on the temporal correlations of a physical observable, measured sequentially at different time intervals during the evolution of a single system. It is well-known that LGI gets violated in the quantum domain serving as a test to validate the notion of macrorealism  in the controversial twilight zone area separating the quantum and classical boundary. 

LGI imposes bounds on the linear combination of two-time correlations between consecutive measurements of a dichotomic  observable of a system respecting macrorealism \cite{emary+14}. As it exhibits a striking symmetry with the Bell-CHSH inequality \cite{MKT+14,DAS+13}, it is also referred to as the temporal Bell inequality \cite{bruckner+04}. Numerous theoretical and experimental investigations have confirmed violations of LGI~\cite{KB07,KB08,Fritz10,WMM10,LMN+10,KB13,UKSR,HSR+13,DBH+11,GAB+11,knee+12,mahesh+11,waldherr+11}.

Akin to the Tsirelson bound for the Bell-CHSH inequality~\cite{Tsirelson80, W06}, investigations towards bounding the temporal correlations for the LGI have been explored~\cite{Fritz10,guehne+13}. Moreover, Budroni and Emary~\cite{BE14} have shown that the strength of violation of the LGIs for \textit{dichotomic} observables can exceed the temporal Tsirelson bound (TTB) in $N$-level quantum systems approaching the \textit{algebraic maximum value} (admitted by the inequality) in the limit of 
$N\rightarrow\infty$. In contrast to the strict adherence to the Tsirelson bound on the Bell-CHSH inequality, violation of the LGI beyond the TTB attracts special attention which sets the tone for the present letter.  
 
 The magnitude of the violation of LGI depends both on the size of the quantum system and the  measurement schemes employed~\cite{BE14}. To this end, stronger violation of the three-term LGI, beyond the TTB, was reported in liquid state nuclear magnetic resonance (NMR) experiment on an ensemble of three level systems by implementing ideal negative result measurements on dichotomic observables~\cite{Katiyar17}. More recently, experimental demonstration of significantly larger violations of three and four term LGI, with three measurement outcomes, have been reported in photonic three-level systems~\cite{WEmaryexpt17}. Possibility of robust violation exceeding the TTB and reaching up to the  algebraic maximum value of the inequality has been theoretically investigated in multiqubit systems by different measurement schemes~\cite{Nori+16} and by constructing variants of LGI other than the standard LGI~\cite{APan+18}.

With the interest in observing violations beyond the temporal Tsirelson bound growing, we explore if it is possible to witness violations of LGI higher than the TTB in a two-level system. To this end, we theoretically investigate the effect of non-unitary evolution generated by a \textit{parity-time}~
($\mathcal{P}\mathcal{T}$) symmetric non-Hermitian Hamiltonian~\cite{BenderPT98} on the possible violations of LGI in a two-level system. The 
$\mathcal{P}\mathcal{T}$ Hamiltonian commutes with the combined symmetry operations of parity 
($\mathcal{P}$), and
time reversal ($\mathcal{T}$); it has real eigenvalues and shares common eigenvectors with 
$\mathcal{P}\mathcal{T}$ operation up to  the exceptional point  or the $\mathcal{P}\mathcal{T}$ symmetry breaking  point. Spontaneous $\mathcal{P}\mathcal{T}$ symmetry breaking occurs at the exceptional point, separating the broken and unbroken phases of $\mathcal{P}\mathcal{T}$ symmetry. The Hamiltonian and the $\mathcal{P}\mathcal{T}$ symmetry operator do not share common eigen states in the regime of broken  
$\mathcal{P}\mathcal{T}$ symmetry, even though they commute with each other~\cite{BenderPT98,BenderBMPT99,BenderBJ02,B05}. Peculiar features exhibited by $\mathcal{P}\mathcal{T}$ symmetric systems (see 
\cite{L17,CY18} for a recent review on theoretical and experimental investigations),  close to the exceptional point, have attracted attention of researchers over the last two decades~\cite{BenderPT98,BenderBMPT99,BenderBJ02,B05,RKE+10,LRCPT,BandwidthPT,rklee+14,Ueda17,Nori+18,Ueda18,Wu+19}. 
Simulating the ever gnawing energy exchange interaction of a quantum system with its environment in terms of  $\mathcal{P}\mathcal{T}$ symmetric non-unitary dynamics offers a new perspective  for the study of open quantum systems. There is  a surge of activity in  experimentally simulating    
$\mathcal{P}\mathcal{T}$ symmetric quantum  dynamics~\cite{Nori+18,Ueda18,Wu+19}. From the information theoretic perspective, in particular, it was shown that the information flow from the system to its environment can be completely retrieved in the unbroken regime of 
$\mathcal{P}\mathcal{T}$ symmetry and vice versa~\cite{Ueda17}. Such reversible-irreversible criticality of information flow under $\mathcal{P}\mathcal{T}$ symmetric dynamics has recently been demonstrated experimentally, where non-unitary quantum dynamics of a two-state system is simulated in a single-photon interferometric network~\cite{Ueda18}. This unidirectional information flow behavior, in the unbroken phase of $\mathcal{P}\mathcal{T}$, suggests applications of  $\mathcal{P}\mathcal{T}$ systems in quantum control. Furthermore, the paradigmatic features characterizing $\mathcal{P}\mathcal{T}$ symmetric systems open up avenues for their potential applicability in quantum information processing tasks.

In this letter, we report violation of the LGI higher than the temporal Tsirelson bound in a two-level system undergoing $\mathcal{P}\mathcal{T}$ symmetric non-unitary dynamics. We show that the magnitude of the violation attains the algebraic maximum value in the limit of the spontaneous breaking of 
$\mathcal{P}\mathcal{T}$ symmetry. While the excess violation of the three term LGI were reported for dimensions larger than 2, it was not known whether the same could be observed in a two-level system. Here we report for the first time that the violation of three term LGI beyond the TTB reaching up to the algebraic maximum value is indeed possible with $\mathcal{P}\mathcal{T}$ symmetry.\\   
\textit{Three-term LGI.\textemdash} We consider the three-term LGI: 
\begin{eqnarray}
-3\leq K_3 &\equiv& C_{21} + C_{32} - C_{31} \leq 1 
\label{eq:lgi3}
\end{eqnarray}
where $C_{ji}=\braket{Q(t_j)\,Q(t_i)}\equiv \braket{Q_j\,Q_i}$ 
denote temporal correlations of a dichotomic observable $Q$ with  outcomes $q=\pm 1$, measured at two different time intervals $t_j>t_i$.

According to the notion of macrorealism, there exists  definite values for the observable $Q$ implying that the outcome of measurement made at an instant of time $t_j$ is not influenced by that made at an earlier time instant $t_i$. This leads to the bound of Eq.(\ref{eq:lgi3}). On the other hand, with appropriate choice of measurement time intervals, it is possible for the Leggett-Garg parameter $K_3$ to exceed the bound 1 (see Eq.(\ref{eq:lgi3})) in quantum systems. In particular, the maximum value of 
$K_3$ for a two-level quantum system undergoing unitary dynamics, when projective measurements 
$\Pi_{i}$ ($\Pi_j$) of an observable $Q$ with outcomes $q_i$ ($q_j$) at time instants $t_i$ ($t_j$) is given by $K_3^{\rm TTB}=3/2$~\cite{LG85}.

Within the standard quantum theoretic framework, for a system evolving under coherent unitary dynamics, violation of the three-term LGI beyond the value $K_3^{\rm TTB}$ could be witnessed by extending the dimension of the Hilbert space (i.e., by considering quantum systems with number of levels $N>2$) and by employing a general measurement update rule~\cite{BE14}. Under this context, it is of interest to verify whether $\mathcal{P}\mathcal{T}$-symmetric dynamics leads to violation of LGI larger than 
$K_3^{\rm TTB}$ in two-level systems. 
\begin{figure}[h!]
	\centering
	\includegraphics[width=0.9\linewidth]{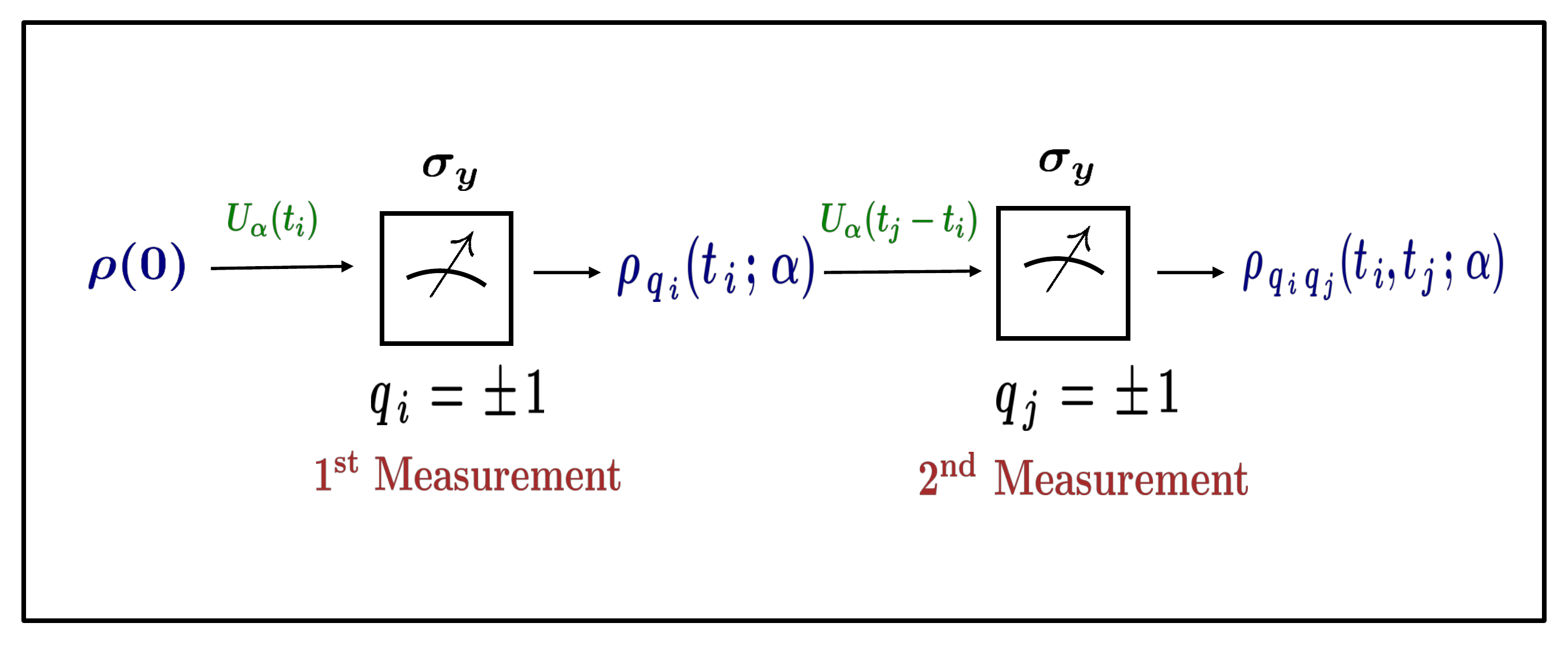}
	\caption{Leggett-Garg inequality two-time measurement scheme.}
	\label{fig:scheme}
\end{figure}

\textit{$\mathcal{P}\mathcal{T}$ symmetric dynamics of a two-level system.\textemdash} 
We consider evolution of a two-level quantum system generated by  the  $\mathcal{P}\mathcal{T}$ symmetric Hamiltonian:
\begin{equation}
H = s 
\begin{pmatrix} 
i\sin{\alpha} & 1 \\
1 &  -i\sin{\alpha} 
\end{pmatrix}    
\hspace{2mm}s,\alpha \in \mathbb{R}; s\neq 0.
\label{eq:ptham}
\end{equation}

The Hamiltonian (\ref{eq:ptham})  commutes with the $\mathcal{P}\mathcal{T}$ operator i.e., 
$[H, \mathcal{P}\mathcal{T}]=0$. It is Hermitian for $\alpha=0$. Though the Hamiltonian is non-Hermitian for $\alpha \neq 0$, it possesses real eigenvalues (by virtue of $\mathcal{P}\mathcal{T}$ symmetry) given by $E_{\pm}=\pm s \cos{\alpha}$, with the corresponding non-orthogonal eigenvectors 
{\small$$\ket{E_{+}(\alpha)} = \frac{e^{i\alpha/2}}{\sqrt{2\cos{\alpha}}}	\begin{pmatrix} 
	1 \\
	e^{-i\alpha} 
	\end{pmatrix}, \hspace{0.5 mm}
	\ket{E_{-}(\alpha)} = \frac{i\, e^{-i\alpha/2}}{\sqrt{2\cos{\alpha}}} \begin{pmatrix} 
	1 \\
	-e^{i\alpha}. 
	\end{pmatrix}$$} 
\noindent Both $H$ and $\mathcal{P}\mathcal{T}$ share common eigenvectors in the range 
$0\leq\alpha<\frac{\pi}{2}$. The eigenvalues and eigenvectors coalesce at the exceptional point  
$\alpha~=~\pi/2$. Here, we restrict ourselves to the parameter range $0\leq \alpha<\pi/2$,  i.e., to the  unbroken $\mathcal{P}\mathcal{T}$ symmetry regime.

The non-unitary time evolution generated by the $\mathcal{P}\mathcal{T}$ symmetric Hamiltonian 
(\ref{eq:ptham}) has the following explicit form: 
\begin{eqnarray}
U_\alpha(t) &=& e^{-i\,t\, H}= \frac{1}{\cos{\alpha}}
\begin{pmatrix} 
\cos{(t^\prime - \alpha)} & - \di \sin{t'} \\
- \di \sin{t'} &  \cos{(t^\prime + \alpha)} 
\end{pmatrix} 
\label{eq:ptunitary}
\end{eqnarray}
where $t^\prime = \frac{\Delta E}{2}\,t$ and $\Delta E = E_{+}-E_{-}=2\,s\,\cos\alpha$;  we have set 
$\hbar =1$. Consider a two-level system prepared in a state $\rho(0)$ (at time $t = 0$). The time evolved state $\rho(t;\alpha)$ under the $\mathcal{P}\mathcal{T}$ symmetric Hamiltonian (\ref{eq:ptham}) is given by 
\begin{eqnarray}
\label{eq:ptrho}
\rho(0) \stackrel{U_\alpha(t)}{\longrightarrow} \rho(t;\alpha)=\frac{U_\alpha(t)\rho(0)U_\alpha^\dag(t)}{{\rm Tr}[U_\alpha(t)\rho(0)U_\alpha^\dag(t)]}.
\end{eqnarray}
The normalization term in the denominator of (\ref{eq:ptrho}) arises due to the non-unitary dynamics of the system.

\textit{Temporal correlations in a two-level system undergoing $\mathcal{P}\mathcal{T}$ symmetric dynamics.\textemdash}
We consider two-time correlations of the observable 
$Q = \sigma_y=\begin{pmatrix} 
0 & -i \\
i  &  0 
\end{pmatrix}$ when the system is undergoing $\mathcal{P}\mathcal{T}$ symmetric dynamics (see Eqns. (\ref{eq:ptunitary}) and (\ref{eq:ptrho})). One-dimensional projection operators employed for measuring  dichotomic outcomes  $q=\pm\, 1$ of the observable $\sigma_y$ are denoted by $\Pi_{q}$.

When measurement of $\sigma_y$ results in the outcome $q_i$ at time $t_i$, the initial state transforms as 
$\rho(0)~\rightarrow~ \rho_{q_i}(t_i;\alpha)~=~\frac{\Pi_{q_i}\rho(t_i;\alpha)\Pi_{q_i}}{p(q_i, t_i;\alpha)}$, with probability $p(q_i, t_i;\alpha)={\rm Tr}[\rho(t_i;\alpha)\,\Pi_{q_i}]$ for different choices of the parameter $0\leq \alpha<\pi/2$. Conditioned on the measurement outcome $q_i$ at time $t_i$, a second  measurement of  $\sigma_y$ at a later time  $t_j>t_i$  leads to (see Fig.~(\ref{fig:scheme}))
\begin{eqnarray*}
	\rho_{q_i}(t_i;\alpha)&\rightarrow&\rho_{q_i\,q_j}(t_i,t_j;\alpha) \\  
	&=&\frac{\Pi_{q_j}U_\alpha(t_j-t_i)\rho_{q_i}(t_i;\alpha)U_\alpha^\dag(t_j-t_i)\Pi_{q_j}}{p(q_j,t_j\vert q_i, t_i;\alpha)}
\end{eqnarray*}
where $$p(q_j, t_j\vert q_i,t_i;\alpha)=\frac{{\rm Tr}[U_\alpha(t_j-t_i)\rho_{q_i}(t_i;\alpha)U_\alpha^\dagger(t_j-t_i)\Pi_{q_j}]}{p(q_i,t_i;\alpha)}$$ denotes the probability of getting the measurement outcome $q_j$ at time $t_j$, when the first measurement outcome is registered as $q_i$ at an earlier time $t_i$. 
\begin{figure}[hb!]
	\centering
	\includegraphics[width=0.9\linewidth]{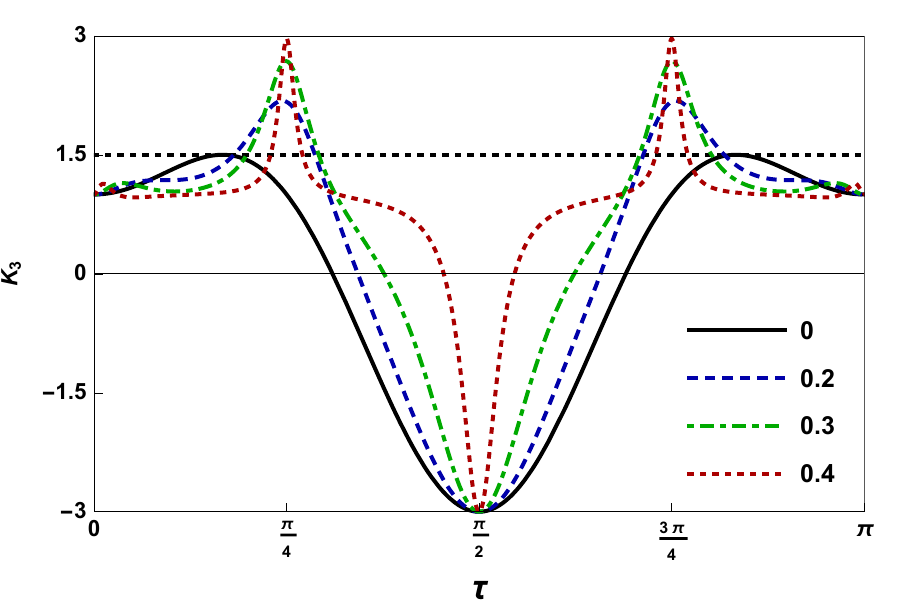}
	\caption{(colour online). Leggett-Garg parameter $K_{3}(\alpha)$ for a two-level system evolving under non-unitary $\mathcal{P}\mathcal{T}$ symmetric Hamiltonian (Eq.(\ref{eq:ptham})) for different choices of $\alpha/\pi$. The (dimensionless) measurement times for measuring the dichotomic observable $\sigma_y$ are set as $t'_1-t'_0=\tau,\, t'_2-t'_1=\tau,\, t'_3-t'_2=\tau,\, t'_3-t'_1=2\,\tau$. Standard three-term LGI corresponding to unitary dynamics is recovered for $\alpha=0$ (solid line); Maximum value of LGI parameter  $K_{3}^{\rm max}(\alpha)$ grows beyond the TTB value $\frac{3}{2}$ achievable  in a two-level system (for $\alpha=0$) with the increase of  $\alpha$ and  approaches its algebraic maximum value 3 in the $\mathcal{P}\mathcal{T}$ symmetry breaking limit $\alpha\rightarrow \pi/2$.}
	\label{fig:k3vst}
\end{figure}

We evaluate the two-time temporal correlations    
$$C_{ji}(\alpha)=\displaystyle\sum_{q_i,q_j=\pm 1}\, q_i\, q_j\, p(q_j,t_j\vert q_i, t_i;\alpha)\, p(q_i,t_i;\alpha)$$      
by considering a two level system initialized to a maximally mixed state $\rho(0)=\frac{I}{2}$, where 
$I$ denotes $2\times 2$ identity matrix, and obtain 
\begin{widetext}
	\begin{eqnarray}
	\label{eq:cji}
	C_{ji}(\alpha)&=&\frac{I_{ji}(\alpha)\left[R_{ji}(\alpha)R_{i0}(\alpha)
		+K_{ji}(\alpha)\, K_{i0}\,\right] 
		+K_{ji}(\alpha)\, \left[R_{ji}(\alpha)\,K_{i0}(\alpha)+R_{i0}(\alpha)\,
		K_{ji}(\alpha)\right]}{\left[R^2_{ji}(\alpha)-K^2_{ji}(\alpha)\right]\,R_{i0}(\alpha)}
	\end{eqnarray}
\end{widetext} where 
\begin{eqnarray*}
	R_{j\,\mu}(\alpha)&=&  1 +\, 2\, \sin^2(t'_j-t'_\mu)\,\tan^2\alpha \\  I_{j\mu}(\alpha)&=&  \cos[2(t'_j-t'_\mu)] - 2\, \sin^2[(t'_j-t'_\mu)\, \tan^2\alpha \\  
	K_{j\mu}(\alpha)&=& 2\,\sin^2[2(t'_j-t'_\mu)]\, \tan\alpha\, \sec\alpha,  \ \ 
\end{eqnarray*}
with $j=1,2,3,$ and  $\mu=0,1,2,\ j>\mu.$\\

\textit{Violation of LGI up to algebraic maximum.\textemdash} 
 Setting  measurement\ \  time intervals 
$t_1'-t'_0~=~\tau,\  t'_2~-~t'_1~=~\tau,  t'_3~-~t_2'~=~\tau$ and $t'_3-t_1'=2\,\tau$,  we compute the left hand side of the three-term LGI  $K_3(\alpha)~=~C_{21}(\alpha)~+~C_{32}(\alpha)~-~C_{31}(\alpha)$.  Standard three-term LGI parameter i.e.,  $K_3=2\,\cos(2\tau)-\, \cos(4\tau)$ is recovered when $\alpha=0$ i.e., when  the dynamics of the system is unitary.

\begin{figure}[h!]
	\includegraphics[width=8.5cm,height=4cm]{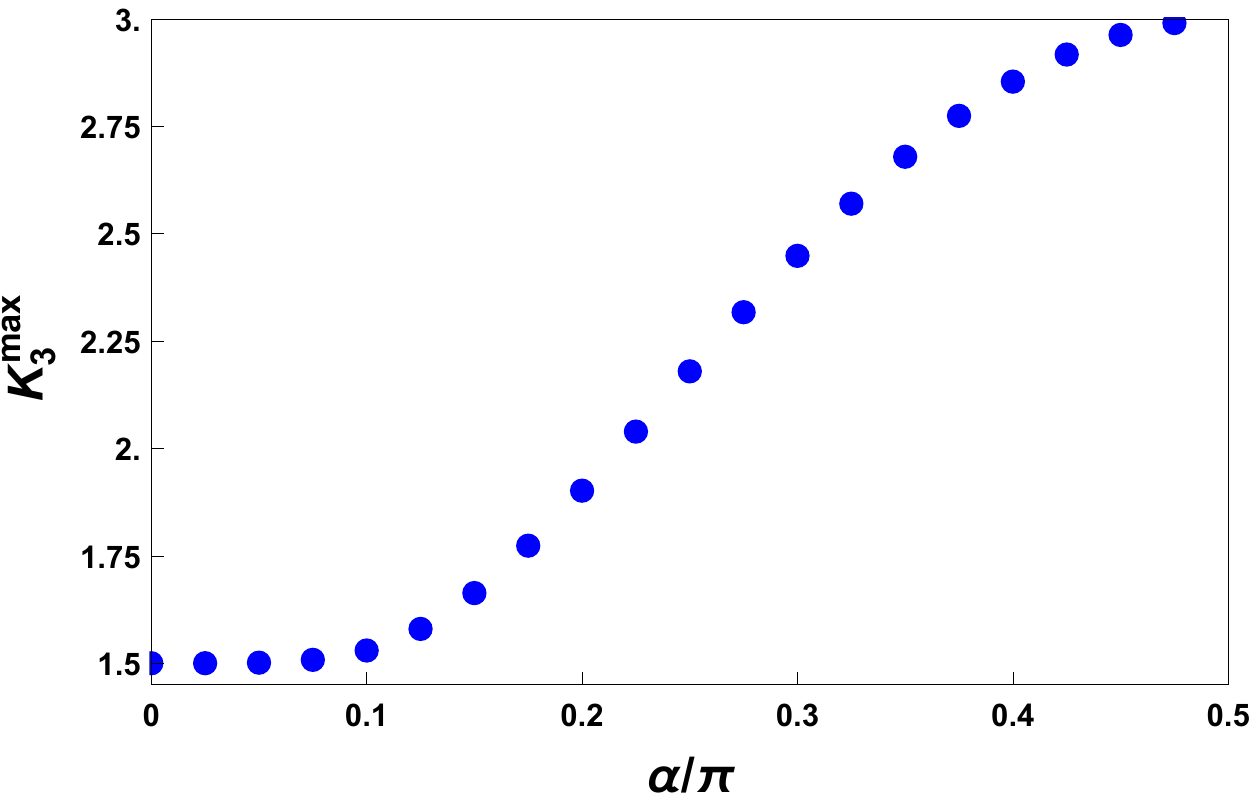}
	\caption{Maximum value $K^{\rm max}_3(\alpha)$ for different choices of  $\alpha/\pi$. Here, the measurement time-step is restricted to $\tau=[0,\pi/4]$ during which the first instance of reaching the maximum value is observed. It may be noted that $K^{\rm max}_3$ approaches the algebraic maximum value 3 of LGI in the limit $\alpha\rightarrow \pi/2$.}
	\label{fig:k3vsalpha}
\end{figure}

\begin{figure}[h]
	\includegraphics[width=4.8cm,height=4.1cm]{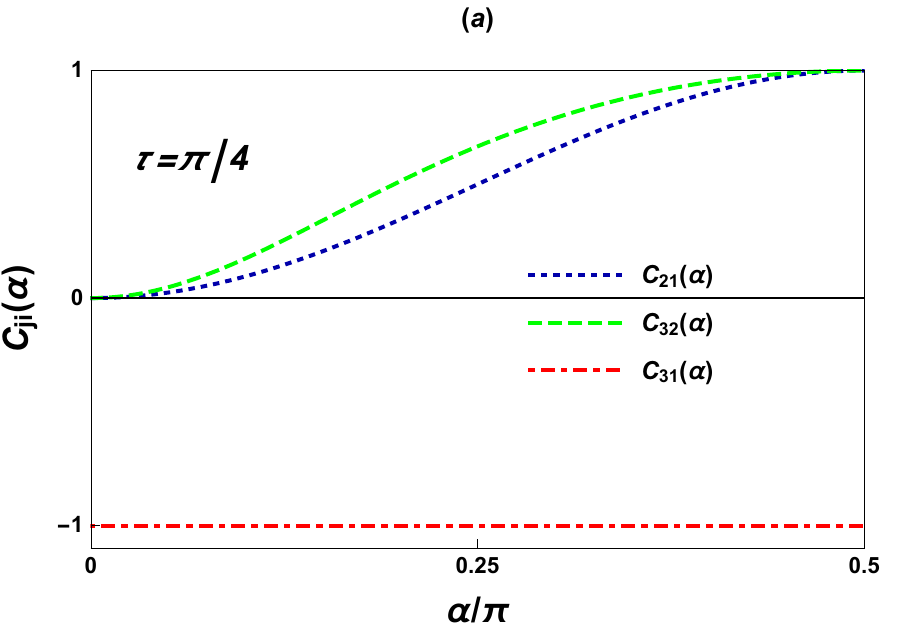}\includegraphics[width=3.5cm,height=4.05cm]{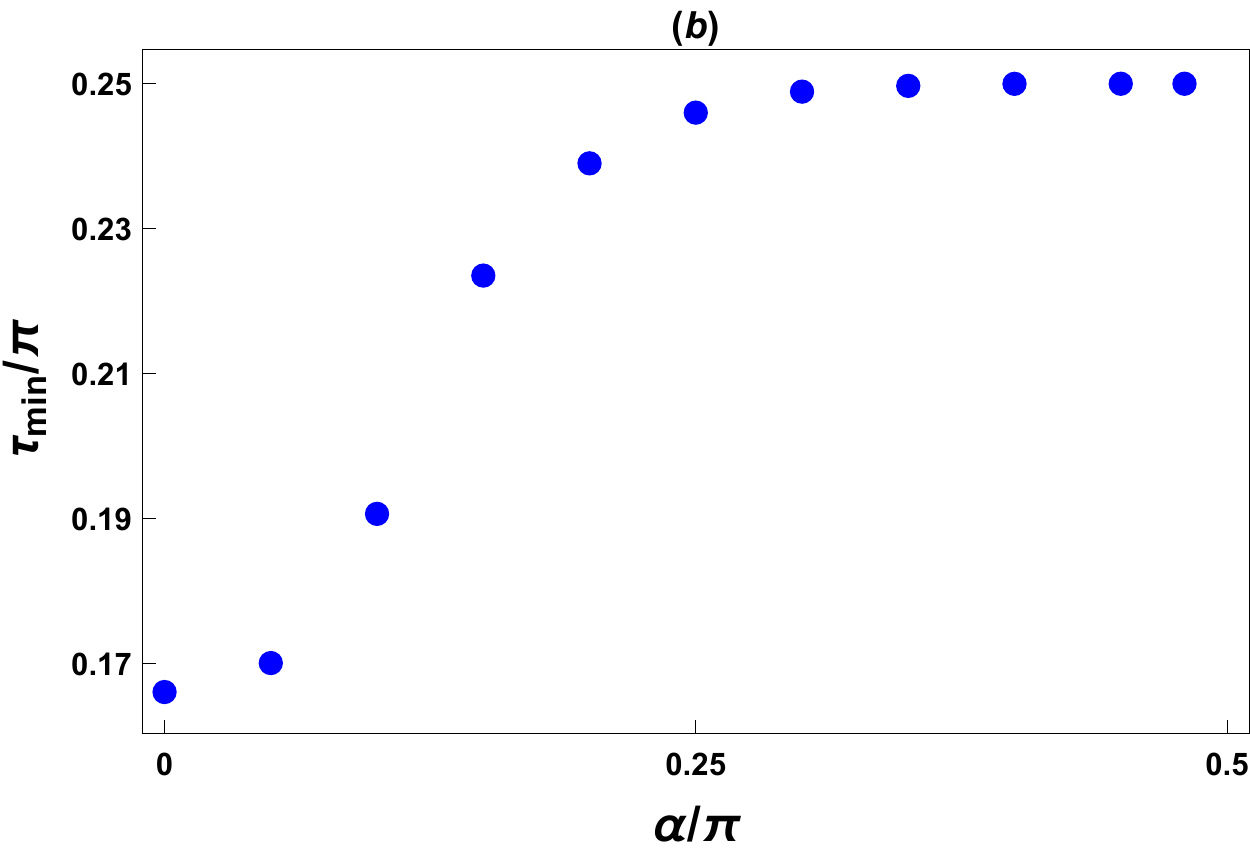}
	\caption{(a) Temporal correlations $C_{21}(\alpha), C_{32}(\alpha)$ and  $C_{31}(\alpha)$ for different choices of $\alpha/\pi$ at a fixed measurement time-step $\tau=\pi/4$. Note that  $C_{21}(\alpha)$, $C_{32}(\alpha)\rightarrow 1$ in the limit  $\alpha\rightarrow\pi/2$  and $C_{31}(\alpha)=-1$ when $\tau=\pi/4$. (b) Plot showing saturation of the shortest value of the measurement time-step $\tau_{\rm min}$ for which $K_3(\alpha)$ attains its maximum value $K^{\rm max}_3$. It may be seen that $\tau_{\rm min}\rightarrow \pi/4$ as $K^{\rm max}_3$ attains the algebraic maximum value 3 in the limit $\alpha\rightarrow \pi/2$ of the spontaneous $\mathcal{P}\mathcal{T}$ symmetry breaking.}       
	\label{fig:Fig4}
\end{figure}

Having computed the two-time correlations, we have plotted $K_3(\alpha)$ as a function of dimensionless measurement time interval $\tau$ for different values of $\alpha$ in Fig.~(\ref{fig:k3vst}).  It is clearly seen that the three-term LGI is violated  all the way up to the algebraic maximum 3, at the 
$\mathcal{P}\mathcal{T}$ symmetry breaking point $\alpha\rightarrow\pi/2.$ 
Variation of $K^{\rm max}_3(\alpha)$ for different choices of the parameter $\alpha/\pi$ is illustrated in the Fig.~(\ref{fig:k3vsalpha}), where it is clearly seen that $K^{\rm max}_3(\alpha)\rightarrow 3$ when $\alpha\rightarrow\pi/2$. We also observe that the shortest measurement time-step $\tau_{\rm min}$, at which  $K^{\rm max}_3$ reaches the algebraic maximum 3, saturates  to a constant value of $\pi/4$ as $\alpha\rightarrow \pi/2$ (see Figs.~\ref{fig:k3vst}, \ref{fig:Fig4}(b)). In fact, when the measurement time-step is fixed to be $\tau=\pi/4$, the temporal correlations (see Eq. (\ref{eq:cji})) $C_{21}(\alpha)$, $C_{32}(\alpha)$ approach the value 1 as  $\alpha\rightarrow \pi/2$, and $C_{31}(\alpha)=-1$, irrespective of the parameter $\alpha$ (see Fig.~(\ref{fig:Fig4})(a)). 

\textit{Discussion.\textemdash} Negation of macrorealism in a two-level system is no surprise as it is genuinely quantum in nature. Nevertheless, the importance of tests of LGI on two-level systems elucidates the persistence and stability of the quantumness of the system in the presence of environmental interactions~\cite{mahesh+11}. This merits investigations on the violation of LGI in a two-level system under a more general  framework of non-Hermitian, non-unital dynamics, thereby tapping the counter intuitive traits, which come underpinned for  applications in future quantum technology. This is for the first instance, to the best of our knowledge, that investigation of macrorealism on a 
$\mathcal{P}\mathcal{T}$ symmetric system is carried out. We have shown here that violation of LGI beyond the TTB and up to the algebraic maximum is possible in a two-level $\mathcal{P}\mathcal{T}$ system. Our work reveals the non-trivial influence of $\mathcal{P}\mathcal{T}$ symmetric dynamics on the temporal correlations of an observable in a two level quantum system.  Significantly, non-unitary dynamics of two-level quantum system under $\mathcal{P}\mathcal{T}$ symmetric Hamiltonians, both in the unbroken and broken regimes, has been simulated experimentally in single-photon interferometric networks~\cite{Ueda18} and also in a single nitrogen-vacancy (NV) center in diamond~\cite{Wu+19}. We believe that our theoretical result on the violation of LGI beyond the TTB, which is shown to be possible in  
$\mathcal{P}\mathcal{T}$ symmetric two-level systems, could be readily verified by the existing experimental platforms~\cite{Ueda18,Wu+19}.        
Furthermore, understanding the basic principle resulting in the algebraic maximum violation of LGI at the $\mathcal{P}\mathcal{T}$ symmetry breaking point deserves a deeper analysis as this may shine light on the possible information content of  quantum temporal correlations in  two-level 
$\mathcal{P}\mathcal{T}$ symmetric systems. Finally, we believe that our work opens up avenues for further studies on $\mathcal{P}\mathcal{T}$ symmetric quantum systems, both from the foundational view point, as well as from the perspective of applicability in quantum information processing.\\  

The authors gratefully acknowledge discussions with Professors A K Rajagopal, Sudha, Sibasish Ghosh. HSK and ASH thank Prof. R Srikanth for insightful comments on the manuscript. ASH is grateful to Prof. C M Chandrashekhar for hosting her visit at IMSc, Chennai during which this work was initiated and partly completed.

\bibliographystyle{apsrev4-1}
\bibliography{LGIPT}
\end{document}